\documentclass[%
aip,
cha,%
amsmath,amssymb,
reprint,%
]{revtex4-1}
\usepackage{color}
\usepackage{graphicx}
\usepackage{dcolumn}
\usepackage{bm}
\usepackage{amsfonts}
\usepackage{sidecap}
\usepackage{float}
\usepackage{parskip}
\usepackage{grffile}

\usepackage{color}
\usepackage{hyperref}
\usepackage{hhline}
\usepackage{mathtools}
\usepackage[normalem]{ulem}
\makeatletter
\def\@eqnnum{{\normalsize \normalcolor (\theequation)}}
\makeatother
\begin{document}
	
	\title{Explosive synchronization in interlayer phase-shifted Kuramoto oscillators on multiplex networks}
	
	\author{Anil Kumar}
	\affiliation{Complex Systems Lab, Discipline of Physics, Indian Institute of Technology Indore, Khandwa Road, Simrol, Indore, 453552, India}
	
	\author{Sarika Jalan}
	\email{sarikajalan9@gmail.com}
	\affiliation{Complex Systems Lab, Discipline of Physics, Indian Institute of Technology Indore,  Khandwa Road, Simrol, Indore, 453552, India}
	
	\affiliation{Centre for Bio-Science and Bio-Medical Engineering, Indian Institute of Technology Indore, Khandwa Road, Simrol, Indore, 453552, India}
	\affiliation{Center for Theoretical Physics of Complex Systems, Institute for Basic Science (IBS), Daejeon, 34126, Korea}

	\date{\today}
	
	\begin{abstract}
		We show that an introduction of a phase parameter ($\alpha$), with $0 \le \alpha \le \pi/2$, in the interlayer coupling terms of multiplex networks of Kuramoto oscillators can induce explosive synchronization (ES) in the multiplexed layers. Along with the $\alpha$ values, the hysteresis width is determined by the interlayer coupling strength and the frequency mismatch between the mirror (inter-connected) nodes. A mean-field analysis is performed to support the numerical results. Similar to the earlier works, we find that the suppression of synchronization is accountable for the origin of ES. The robustness of ES against changes in the network topology and frequency distribution is tested.  
		Finally, taking a suggestion from the synchronized state of the multiplex networks, we extend the results to the classical concept of the single-layer networks in which some specific links are assigned a phase-shifted coupling.
		Different methods have been introduced in the past years to incite ES in coupled oscillators; our results indicate that a phase-shifted coupling can also be one such method to achieve ES.        
	\end{abstract}
	
	\maketitle
	
	\begin{quotation}
A first-order phase transition or explosive synchronization (ES) is an abrupt transition from an incoherent state to a coherent state or vice versa. \cite{Boccaletti2016, Raisa2019} 
Furthermore, due to their relevance to real-world systems, the phase-shifted Kuramoto oscillators have been studied extensively \cite{Lohe2015} and therefore it is worthy to investigate if a phase-shift can induce ES in these oscillators. Here we show that a phase shift in the interlayer coupling terms of a multiplex network of the Kuramoto oscillators can trigger ES in the layers. The multiplex networks, a special form of multilayer networks, have drawn considerable attention in recent years. \cite{DeDomenico2013, Boccaletti2014, Kivela2014} 
They are useful to characterize complex systems having different types of interactions among the nodes. We extend the results to the single-layer networks in which some specific links have phase-shifted interactions.  The interlayer phase-shifted model may be relevant to real-world systems.  For example, the neural dynamics represented by Wilson-Cowan oscillators can be reduced to a multilayer network of phase oscillators having phase-shifted interlayer couplings. \cite{Maximilian2015, Bastian2019}
	\end{quotation}
	
	\section{Introduction} 
Different methods have been proposed in the past years to incite ES in the single-layer networks of first-order Kuramoto oscillators. Degree frequency correlation, \cite{Gardenes2011} 
frequency-weighted coupling, \cite{Zhang2013,Leyva2013b} adaptive coupling, \cite{Zhang2015} 
disorder in natural frequencies, \cite{Skardal2016} and repulsive coupling \cite{Hyunsuk2011} etc. are some of the examples. 
Most of these works have found that suppression of synchronization is the key mechanism behind the origin of ES. A few experiments have also been performed to demonstrate the presence of ES; for instance, ES in mercury-beating heart oscillators and chaotic oscillators. \cite{PKumar2015, Leyva2015} 
	Furthermore, multiplex networks, a special form of multilayer networks, have drawn considerable attention in recent years. \cite{DeDomenico2013, Boccaletti2014, Kivela2014} In this framework the same set of nodes are replicated in different layers; also, every node in a layer is connected to its mirror node across the layers.  The multiplex framework is useful to characterize those complex systems where different types of interactions or connections can exist. Some of the most common examples are from social, transport, and neural networks. \cite{Boccaletti2014}
	The results from such a setup may differ considerably from their single-layer counterparts.   
	A few works have displayed ES in coupled oscillators on multiplex or multilayer networks. For example, ES in intra or interlayer adaptively coupled oscillators, \cite{Danziger2019, Kumar2020, Khanra2018, Kachhvah2020} frequency mismatched oscillators, \cite{Jalan2019} repulsively coupled oscillators, \cite{Vasundhara2019} time-delayed oscillators, \cite{Ajaydeep2019} and higher-order coupled oscillators. \cite{Skardal2019} 
  
	Although the phase-shifted coupling in Kuramoto oscillators was introduced a long time back by Sakaguchi and Kuramoto, a simple set up of uniformly phase-shifted oscillators on the single-layer networks does not exhibit any ES. \cite{Sakaguchi1986} Similarly, to our knowledge there has not been any investigation manifesting the emergence of ES due to distributed phase shifts in the Kuramoto oscillators. \cite{Lohe2015} Although some studies have shown ES in the phase-shifted oscillators, the underlying cause is not the phase-shift. \cite{Xiao2017, Khanra2020} 
	Another model very similar to the multilayer networks, the multiple populations of identical and non-identical phase oscillators, have shown existance of different phenomena such as chimera, chaos due to a phase shift in the intra or interlayer couplings. \cite{Abrams2008, Laing2009, Barreto2008, Bick2018, Montbrio2004, Zhang2016} However, ES in the populations due to a phase-shifted coupling has not been demonstrated. 
	Notably, few works have taken the same model as ours but with identical oscillators; they have shown the presence of chimera states due to a phase shift in the intra or interlayer couplings. \cite{Maksimenko2016, Frolov2018}
	
In this work, we show that a phase shift in the interlayer coupling terms of a multiplex network of the Kuramoto oscillators can trigger ES in the layers. Sec. \ref{sec_3} emphasises that as $\alpha \rightarrow \pi/2$, the ES emerges along with a hysteresis. Along with $\alpha$ values, the hysteresis width depends on the interlayer coupling strength and the frequency mismatch between the mirror nodes. 
	Sec. \ref{sec_4} tests the robustness of ES against changes in the network topology and frequency distribution.
	A mean-field analysis is performed in Sec.\ref{sec_5} to support the numerical results. 
	Finally, for a particular set up of natural frequencies, in Sec.\ref{sec_6} we show that a phase-shifted model can trigger ES even in the single-layer networks and therefore extend the results to the basic framework of the networks.
	
	\section{Model}
	\label{section_model}
	We take a multiplex network of two globally connected layers; each node in a layer is represented by a Kuramoto oscillator. \cite{Kuramoto_1984} An $i^{th}$ node in layer $a(b)$ is connected with its mirror node in layer $b(a)$ through an interlayer coupling having a phase shift $\alpha$. Therefore, the angular velocity of an $i^{th}$ node in layer $a(b)$ is given by
	\begin{subequations}
		\begin{align}
			\dot{\theta^{a}_i} = \omega_i^{a} + \frac{\sigma}{N} \sum_{j=1}^N \sin(\theta_j^{a}-\theta_i^{a})+
			\lambda \sin(\theta_i^{b}-\theta_i^{a} + \alpha), \tag{1-a}\label{eq1} \\
			\dot{\theta_i^{b}} = \omega_i^{b} + \frac{\sigma}{N} \sum_{j=1}^N \sin(\theta_j^{b}-\theta_i^{b})+
			\lambda \sin(\theta_i^{a}-\theta_i^{b}+\alpha), \tag{1-b}\label{eq2}
		\end{align}\label{eq_model} 
	\end{subequations}
	where $i=1,2,3, . . .,N$. $\theta_i^{a(b)}$ and $\omega_i^{a(b)}$ represent phase and natural frequency of the $i^{th}$ node in layer $a(b)$. Initial phases of the oscillators are drawn from a uniform random distributions in the range $-\pi \le \theta_{i}^{a(b)} \le \pi$. To keep the natural frequency distribution same for both the layers, they are taken from the relation $\omega_{i}^{a(b)} =-0.5+(i-1)/(N-1)$, where $i=1, 2, . . .,N$. However, the mirror nodes have different natural frequencies in general i.e. $\omega_i^a \neq \omega_i^b$. $\sigma$ denotes intra-layer coupling strength among the nodes of layer $a(b)$,
	whereas $\lambda$ represents the inter-layer coupling strength. In the entire work, except Fig.\ref{fig_hysteresis_size}(a-c), ES is generated by varying the phase shift $\alpha$ from $0$ to $\pi/2$. 
	Eq.\ref{eq_model} is solved numerically using Runge-Kutta $4^{th}$ order method with adaptive time step size. \cite{rk4_method}
	The nature of the phase transition is identified by plotting $\sigma$ versus time averaged phase order parameter $r^a$; it is defined as 
	\begin{equation}
		r^{a(b)} e^{i\psi^{a(b)}} = \frac {1}{N} \sum_{j=1}^N  e^{i \theta^{a(b)}_{j}},
		\label{eq_r_a_b}
	\end{equation}
	where $0 \le r^{a(b)} \le 1$. The minimum value, $r^{a(b)}=0$, corresponds to a uniform distribution of the oscillators over a unit circle, while the maximum value, $r^{a(b)}=1$, corresponds to the exact phase synchronization. Time average of $r^{a(b)}$ is taken for $1000$ time steps after neglecting the initial $1000$ time steps. 
	
	\begin{figure}[t]
		\centerline{\includegraphics[width=\columnwidth]{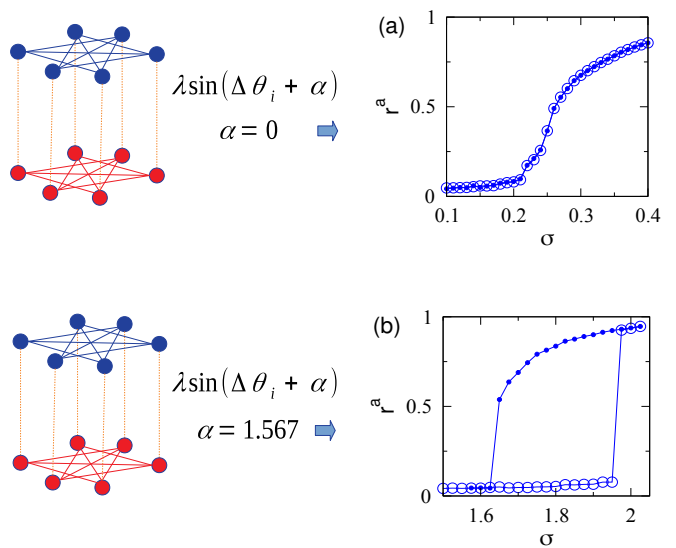}}
		\caption{(color online). (a), (b) $r^a$ as a function of $\sigma$, showing that an increase in $\alpha$ turns a continuous phase transition into a discontinuous one. The schematic diagrams of the multiplex network on the left are shown only for presentation.
			The layers are globally connected networks with $N=1000$, $\lambda =10$, and $\Delta \omega \approx 0.8$. 
			From here onwards, unless mentioned otherwise, the lines with open and filled circles correspond to the forward and backward continuation of $\sigma$, respectively.} 
		\label{fig_multiplexing_effect}
	\end{figure}

	\section{Interlayer phase shift gives rise to ES}
	\label{sec_3}
	We find that an introduction of a phase parameter ($\alpha$) in the interlayer coupling terms of a multiplex network drives ES in layer $a(b)$. At $\alpha=0$, Fig. \ref{fig_multiplexing_effect}(a) illustrates that $r^a$ increases continuously with an increase in $\sigma$, while an increase in $\alpha$ to $1.567$ turns the phase transition to a discontinuous one [Fig. \ref{fig_multiplexing_effect}(b)]. To demonstrate if there exist any hysteresis along with the discontinuity, we plot $r^a$ values with respect to decreasing $\sigma$ values. As displayed by Fig. \ref{fig_multiplexing_effect}(b), the discontinuity in ES is accompanied by the hysteresis.
	While increasing or decreasing $\sigma$, the phases at the last time step at a $\sigma$ value are used as initial phases at the next $\sigma$ value.  
	Since the layers are identical in the network parameters i.e. topology, natural frequencies, and intralayer coupling, the nature of the phase transition should also be the same for both the layers; therefore, only $r^a$ values are plotted with respect to $\sigma$. 
	
	\subsection{Hysteresis width}
	Along with $\alpha$ values, we find that the hysteresis width ($\Delta \sigma$) depends on the interlayer coupling strength ($\lambda$), and a frequency mismatch between the mirror nodes ($\Delta \omega$).
	The parameter $\Delta \omega$ is defined as
	\[ \Delta \omega = \frac{\sum_{i=1}^N \lvert {\omega_i^a -\omega_i^b} \rvert }{2 \sum_{i=1}^N \lvert{\omega_i^a}\rvert }, \]
	where $0 \le \Delta \omega \le 1$. 
	It is a measure of the total distance between the mirror node's natural frequencies. The minimum value corresponds to $\omega_i^a  = \omega_i^b $ (an identical distribution of the natural frequencies), while the maximum value corresponds to $\omega_i^a  = - \omega_i^b$. We mention that other random configurations of natural frequencies may also satisfy the relation $\Delta \omega = 1$ but we only take $\omega_i^a  = - \omega_i^b$.
	A desired $\Delta \omega$ value is achieved as follows. Starting with $\Delta \omega=0$, two pairs of mirror nodes are chosen randomly and the natural frequencies of the selected nodes in layer $b$ are swapped. The change is accepted if the later value of $\Delta \omega$ is closer to the desired $\Delta \omega$ value; otherwise, the change is rejected. This process is repeated until a desired $\Delta \omega$ value is obtained.
	
\begin{figure}[t]
		\centering
		\centerline{\includegraphics[width=1\columnwidth]{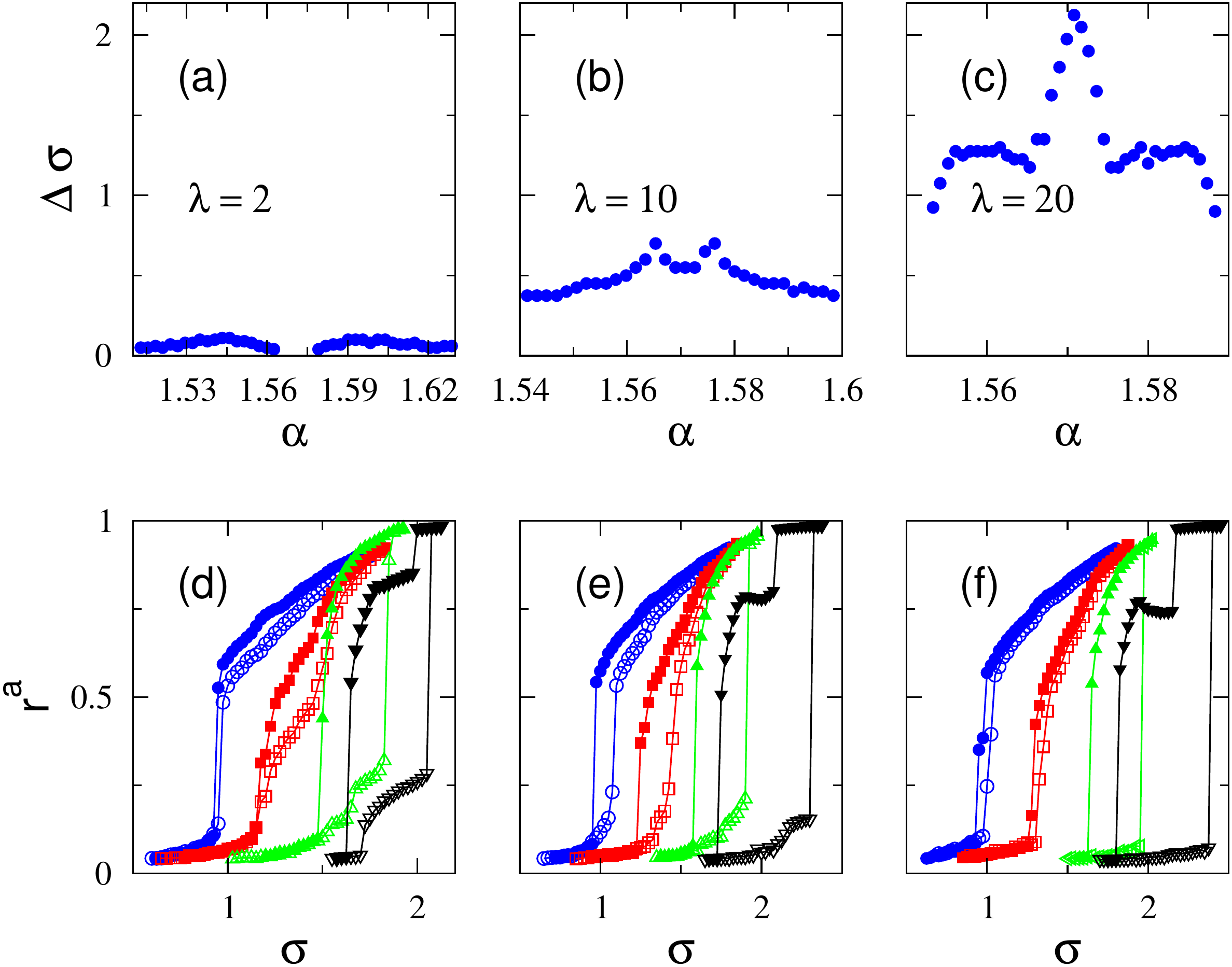}}
		\caption{(color online). (a-c) $\Delta \sigma$ versus $\alpha$ displays a favourable impact of $\lambda$ on the hysteresis width. Here $\Delta \omega$ is fixed at $1$. (d-f) $r^a$ vs $\sigma$ shows the dependence of ES or hysteresis width on $\Delta \omega$. 
			Here (d-f) represents the plots for $\alpha \approx 1.5561, 1.5634, 1.5671$, respectively; in each figure, the lines with circles, squares, upper-triangles, and lower-triangles correspond to $\Delta \omega \approx 0.4, 0.6, 0.8, 1$, respectively. In all the figures, the layers are globally connected with $N= 1000$.} 
		\label{fig_hysteresis_size}
	\end{figure}

	At $\Delta \omega=1$ i.e. at $\omega_i^a=-\omega_i^b$, Figs. \ref{fig_hysteresis_size}(a-c) exhibit that an increase in $\lambda$ results in a larger hysteresis width. This observation is expected because if the phase-shifted coupling term is responsible for ES, an increase in its magnitude should have a favorable impact on ES.  
	Interestingly, the hysteresis width does not increases monotonically as we approach towards $\alpha=\pi/2$. For smaller $\lambda$ values the hysteresis width is maximum if $0 < \alpha < \pi/2$ or $\pi/2< \alpha <\pi$ [Fig. \ref{fig_hysteresis_size}(a)], while for large $\lambda$ values it is maximum at $\alpha=\pi/2$ [Fig. \ref{fig_hysteresis_size}(c)]. 
	Since $\Delta \sigma$ values are symmetric around $\alpha=\pi/2$; therefore, for a better view of the plots here we present the results for $0<\alpha<\pi$; the rest of the paper presents all the results for $0 \le \alpha \le \pi/2$ only.
	Another parameter which affects the ES or hysteresis width is $\Delta \omega$. Figs. \ref{fig_hysteresis_size}(b-e) show that for a given $\lambda$ and $\alpha$ values, an increase in $\Delta \omega$ changes an almost continuous phase transition to a discontinuous one. Using a mean-field analysis, later we justify the impact of $\Delta \omega$ on ES (see Appendix \ref{appendix_a}).  \\
	
	We find that jump size in the backward continuation of $\sigma$ is not significant in some numerical simulations; therefore, there is no critical coupling at which a discontinuity in the backward transition occurs. To follow a consistent approach to measure the hysteresis width, $\Delta \sigma$ in Figs.\ref{fig_hysteresis_size}(a-c) represent the difference between the couplings corresponding to the meeting points of $r$ values in forward and backward direction.   
	Moreover, only $\Delta \sigma$ values close to $\alpha= \pi/2$ are plotted in Figs. \ref{fig_hysteresis_size}(a-c) because we do not find ES for other $\alpha$ values in the range $0$ to $\pi$. As discussed later in Fig. \ref{fig_analytical}(c), although for some $\alpha$ values there can exist a bistable regime without any jump. Since such a hysteresis is not associated with ES, it is neglected in Figs \ref{fig_hysteresis_size}(a-c). Similarly, it is also possible that along with a hysteresis consisting of a significant jump another hysteresis without a significant jump can also exist [Fig.\ref{fig_analytical}(d)], which is also neglected in Figs. \ref{fig_hysteresis_size}(a-c).  
	
	\begin{figure}[t]
		\centering
		\centerline{\includegraphics[width=1\columnwidth]{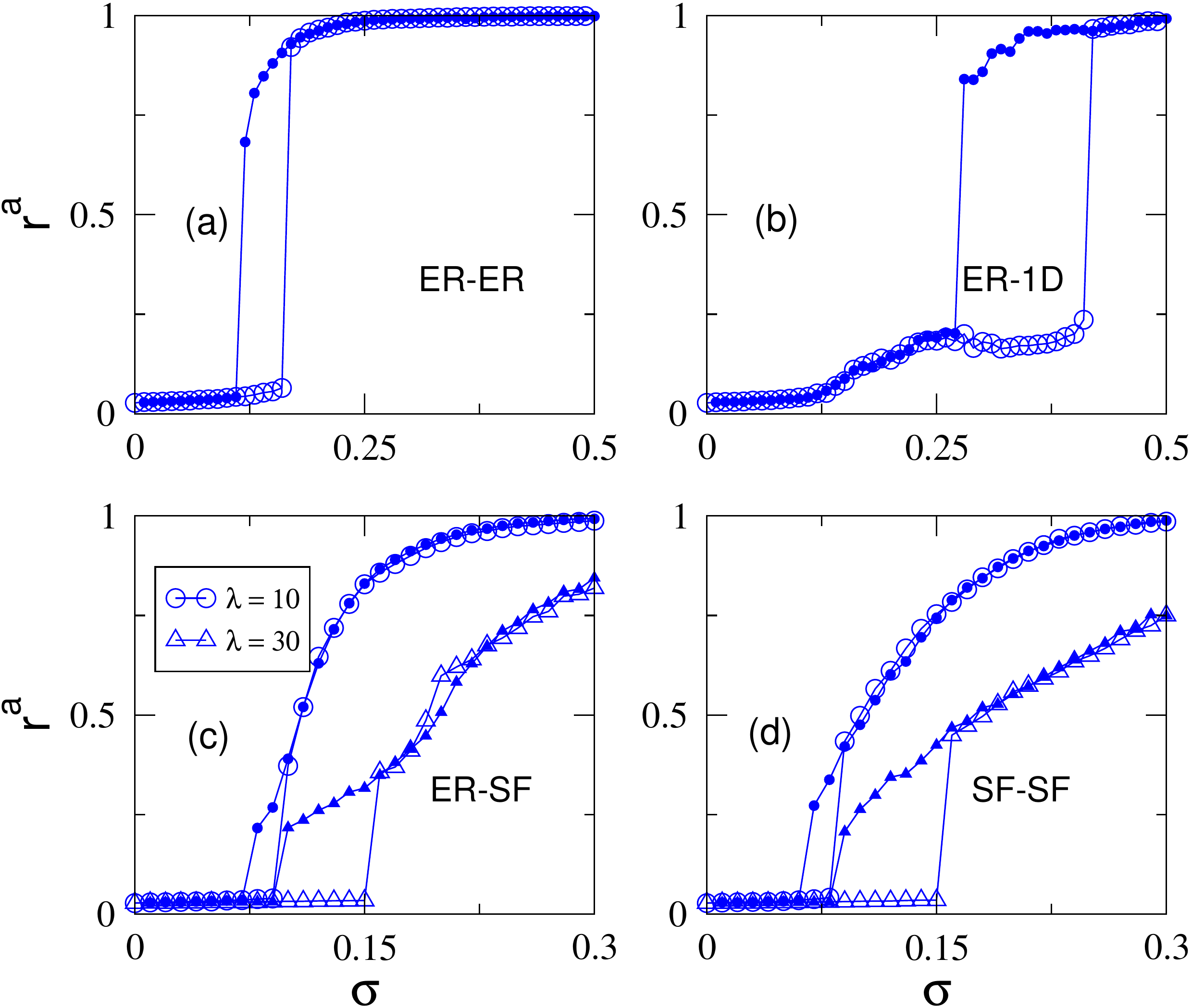}}
		\caption{(color online). $r^a$ versus $\sigma$ in (a-d) display the effect of change in network topology on ES. The network parameters are $N=1000$ and $\lambda=10$. The natural frequencies are assigned randomly, which leads to $\Delta \omega \approx 0.67$. The average connectivity ($ <k^{a(b)}> $) of both the layers is $<k^a> =<k^b> = 16$. To reduce the time taken by the numerical simulations, here we have used RK4 method with a fixed step size of $0.01$. 
		} 
		\label{fig_topology_effect}
	\end{figure}
	
	\section{ES in random networks and non-uniform frequency distributions}
	\label{sec_4}
	Besides the globally connected networks, the phase-shifted coupling can induce ES in random networks having a homogenous degree distribution. However, if the degree distribution is heterogeneous, the jump size is reduced significantly. We consider Erd\"os-R\'enyi (ER) random network, \cite{Erdos1959} regular-ring network, \cite{barabasi1999} and Barabasi and
	Albert's scale-free (SF) network \cite{barabasi1999} for the layers.
	In the numerical simulations, intralayer coupling $\frac{\sigma}{N} \sum_{j=1}^N \sin(\theta_j^{a(b)}-\theta_i^{a(b)})$ in Eq. \ref{eq_model} is replaced by $\sigma \sum_{j=1}^N A_{ij}^{a(b)} \sin(\theta_j^{a(b)}-\theta_i^{a(b)})$. \cite{Arenas2008} Here $A_{ij}^{a(b)} =1$ if $i,j^{th}$ nodes are connected in the layer $a(b)$, while it is $0$ otherwise. Figs. \ref{fig_topology_effect}(a-b) depicts that, as found for the globally connected layers, a multiplex network consisting of ER-ER random network and ER-regular network exhibit ES along with a hysteresis. 
	However, the position of the hysteresis and its width may change with a change in the topology.

	Next, keeping all the network parameters the same, we multiplex a homogeneous network (ER network) with a heterogeneous network (SF network) or take both layers as heterogeneous networks.
	For these topologies, although the hysteresis exist but the jump size is decreased significantly [\ref{fig_topology_effect}(c,d)]. 
	Since an increase in $\lambda$ has a favorable impact on ES, we increase the $\lambda$ value from $10$ to $30$ to see if an ES can be generated. However, as illustrated by Figs. \ref{fig_topology_effect}(c),(d) the jump size remains almost same. Therefore, we conclude that heterogeneous networks have an adverse impact on the ES. 
	In contrast to the ER networks, SF networks favor the onset of the synchronized state from an almost zero coupling value. \cite{Arenas2008} 
	Possibly, the interlayer phase shift could not suppress the onset of synchronization in these networks and therefore it failed to induce a significant jump in ES for these topologies.
	
\begin{figure}[t]
		\centering
		\centerline{\includegraphics[width=1\columnwidth]{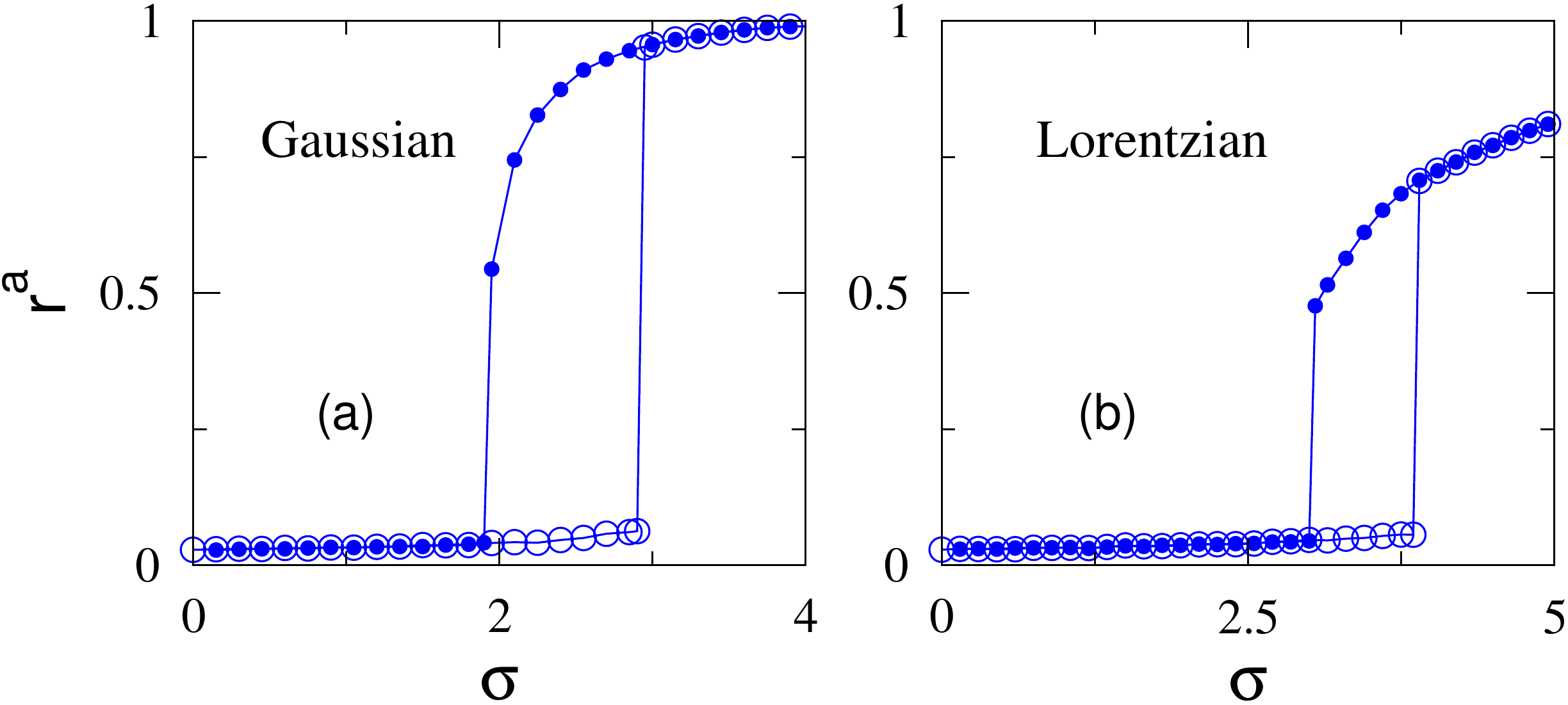}}
		\caption{(color online). The plots of $r^a$ vs $\sigma$ exhibit ES for the non-uniform frequency distributions. The mean of the natural frequencies is $0$ in both the distributions; the standard deviation for the gaussian distribution and the half width at the half maxima for the Lorentzian distribution are both $0.5$. The layers are globally connected with $N=1000$ (for Gaussian) and $N=998$ (for Lorentzian), $\lambda=10$, and $\alpha=\pi/2$. The natural frequencies are assigned randomly.
		} 
		\label{fig_freq_effect}
	\end{figure}

	So far we have shown ES for the uniform distribution of the natural frequencies. Next, we explore the effect of change in the natural frequency distribution on ES.
	Keeping the layers globally connected and natural frequency distribution for the layers identical, we take two important symmetric natural frequency distributions: Gaussian and Lorentzian. Figs. \ref{fig_freq_effect}(a,b) illustrate that the phase transition is discontinuous along with a hysteresis for the considered non-uniform frequency distributions. Therefore, the interlayer phase shift is capable of inducing ES for uniform as well as non-uniform natural frequency distributions. We mention that the Lorentzian distribution is generated using inverse transform sampling. \cite{Lorentz_frequencies} Also, the numerical simulations for the Lorentzian distribution take an unusually long time due to two natural frequencies of the order of $10^{15}$; hence, these natural frequencies are deleted from the distribution. 
	
	\section{Mean field analysis}
	\label{sec_5}
	Assuming that $r^{a(b)}, \psi^{a(b)}$ are constant in time after a steady state is reached, we derive $r^{a(b)}$ values using the well known mean-field approach. \cite{Arenas2008}
	In terms of $r^{a(b)}$, $\psi^{a(b)}$ (Eq.\ref{eq_r_a_b}), Eq.\ref{eq_model} can be rewritten as
	\begin{subequations}
		\begin{align}
			\dot{\theta^{a}_i} = \omega_i^{a} + \sigma r^a \sin(\psi^a-\theta_i^{a})+
			\lambda \sin(\theta_i^{b}-\theta_i^{a} + \alpha),  \\
			\dot{\theta^{b}_i} = \omega_i^{b} + \sigma r^b \sin(\psi^b-\theta_i^{b})+
			\lambda \sin(\theta_i^{a}-\theta_i^{b} + \alpha), 
		\end{align}\label{eq_model_interms_of_r}
	\end{subequations}
	where $i=1, 2, 3, \hdots, N$. Now we assume that $r^a =r^b=r$. As shown later, this is a valid assumption for identical natural frequency distribution in the layers and for high $\lambda$ values.
	Furthermore, an oscillator locked with $\psi^{a(b)}$ rotates with a velocity $\dot{\theta^{a(b)}_i} =\dot \psi^{a(b)} =\Omega$. Note that the time independent intralayer coupling terms suggest that 
	the mirror nodes are locked simultaneously. 
	With these facts,
	Eq. \ref{eq_model_interms_of_r} for a locked oscillator becomes
	
	\begin{subequations}
		\begin{align}
			\Omega &= \omega_i^{a} + \sigma r \sin(\psi^a-\theta_i^{a})+
			\lambda \sin(\theta_i^{b}-\theta_i^{a} +\alpha),\\
			\Omega &= \omega_i^{b} + \sigma r \sin(\psi^b-\theta_i^{b})+
			\lambda \sin(\theta_i^{a}-\theta_i^{b} +\alpha).
		\end{align}\label{eq_synchronized_state}
	\end{subequations}
	
	Sum and difference of Eq. \ref{eq_synchronized_state}(a) and Eq. \ref{eq_synchronized_state}(b) leads to  
	
	\begin{subequations}
		\begin{align} 
			2 \Omega &= \omega_i^{a}+\omega_i^{b} + \nonumber \\ 
			&2 \sigma r \sin \big(\frac{\psi^{a}+\psi^{b} -\theta_i^{a}-\theta_i^{b}}{2} \big) \cos\big(\frac{{\psi^{a}-\psi^{b}} + \theta_i^{b}-\theta_i^{a}}{2} \big) \nonumber \\
			&+ 2 \lambda sin (\alpha) \cos (\theta_i^{b}-\theta_i^{a} ), \\
			0  &= \omega_i^{a}-\omega_i^{b} + \nonumber \\
			& 2 \sigma r \cos \big(\frac{\psi^{a}+\psi^{b}- \theta_i^{a}-\theta_i^{b}}{2} \big) \sin \big(\frac{{\psi^{a}-\psi^{b}} + \theta_i^{b}-\theta_i^{a}}{2} \big) \nonumber \\
			&+ 2 \lambda cos(\alpha) \sin (\theta_i^{b}-\theta_i^{a} ).
		\end{align}
		\label{eq_sum_and_difference}
	\end{subequations}
	
	After putting $\cos\big(\frac{\psi^a+\psi^b - \theta_i^{a}-\theta_i^{b}}{2}\big)$ from Eq.\ref{eq_sum_and_difference}(b) in Eq.\ref{eq_sum_and_difference}(a), we get
	\begin{align}
		&{(\omega_i^{b}-\omega_i^{a} + 2 \lambda \cos(\alpha)\sin (\Delta \theta_i))}^2 \cos^2{(\frac{\Delta \psi -\Delta \theta_i}{2})} + \nonumber\\
		&{(2\Omega -\omega_i^{b}-\omega_i^{a} - 2 \lambda \sin(\alpha) \cos(\Delta \theta_i))}^2 \sin^2{(\frac{\Delta \psi - \Delta \theta_i}{2})} \nonumber\\
		&= \sigma^2 r^2 \sin^2{({\Delta \psi -\Delta \theta_i})},
		\label{eq_theta_a_minus_theta_b}
	\end{align}
	where $\Delta \psi$ and $\Delta \theta_i$ represent $\psi^a-\psi^b$ and $\theta_i^a-\theta_i^b$, respectively.
	As explained in Appendix \ref{appendix_a}, we find the roots of Eq.\ref{eq_theta_a_minus_theta_b} numerically. 
	Next, a comparison of real and imaginary terms of Eq. \ref{eq_r_a_b} leads to  
	\begin{equation}
		r^{a} = \frac {1}{N} \sum_{locked}  \cos(\theta^{a}_{j} -\psi^a) + \frac {1}{N} \sum_{drifting}  \cos(\theta^{a}_{j} -\psi^a), 
		\label{eq_r_a_with_real_part}
	\end{equation}
	
	\begin{equation}
		\sum_{locked}  \sin(\theta^{a}_{j} -\psi^a) + \sum_{drifting}  \sin(\theta^{a}_{j} -\psi^a)=0. 
		\label{eq_r_a_with_imaginary_part}
	\end{equation}
	
	Here the locked oscillators are those for which Eq. \ref{eq_theta_a_minus_theta_b} has a root, while the remaining are the drifting oscillators.\\

	For $\Delta \omega=0$, the synchronized state for the multiplex network is the same as for the isolated network, which is already known, and therefore we ignore this case (Appendix \ref{appendix_a}).		
	For $\Delta \omega=1$, Eq.\ref{eq_theta_a_minus_theta_b} has roots for natural frequencies $\pm \omega_i$; hence, the locked oscillators are placed symmetrically around the mean natural frequency.
	With this fact, the drifting oscillators in Eq.\ref{eq_r_a_with_real_part} and Eq.\ref{eq_r_a_with_imaginary_part} can be neglected in the limit $N \rightarrow \infty$. \cite{Kumar2020} For $0<\Delta \omega<1$, a matching of numerical and analytical results (Appendix \ref{appendix_b}) indicate that we can neglect the drifting oscillators.
	Therefore, Eq.\ref{eq_r_a_with_real_part} reduces to
	
	\begin{equation}
		r^{a} = \frac {1}{N} \sum_{locked}  cos(\theta^{a}_{j} -\psi),
		\label{eq_r_a_with_real_part_drifting_osc_neglected}
	\end{equation}
	
	while Eq. \ref{eq_r_a_with_imaginary_part}, with the help of Eq. \ref{eq_synchronized_state}(a), can be written as
	\begin{equation}
		\Omega= \frac{1} {N_l} \sum_{locked} \omega_j^a + \lambda \sin(\theta^{b}_{j} -\theta^{a}_{j} + \alpha). 
		\label{eq_Omega}
	\end{equation}

\begin{figure}[t]
		\centering
		\centerline{\includegraphics[width=\columnwidth]{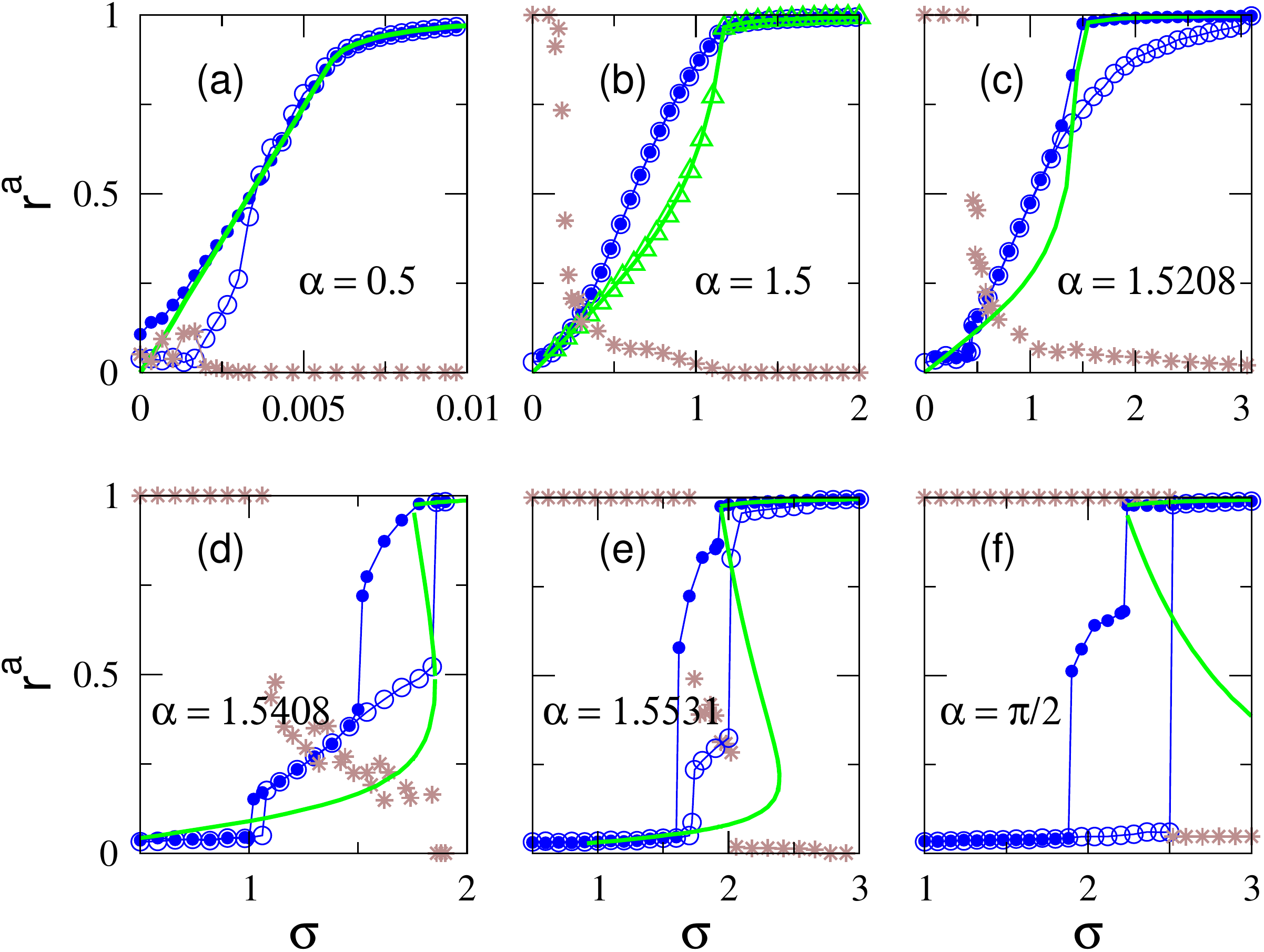}}
		\caption{(color online). The plots in (a-f) compares $r$ values from the mean field analysis (the continuous green lines) with the corresponding $r$ values from the numerical simulations (the lines with symbols). The stars represent maximum fluctuation in $|sin(\psi^a-\psi^b)|$ in the forward continuation of $\sigma$ after the initial transient time is passed. The triangles in (b) represent mean field predicted $r$ values when Eq.\ref{eq_r_a_with_real_part_drifting_osc_neglected} and Eq.\ref{eq_Omega} are solved for $N=1000$.
			The network parameters in the numerical simulations are $N=1000$, $\lambda=10$, and $\Delta \omega =1$.} 
		\label{fig_analytical}
	\end{figure}

	Here $N_l$ represents the total number of locked oscillators. 
	Using Eq.\ref{eq_synchronized_state} (a) and Eq.\ref{eq_theta_a_minus_theta_b}, we find $r, \Omega$ values by solving Eq.\ref{eq_r_a_with_real_part_drifting_osc_neglected} and Eq.\ref{eq_Omega} numerically. 
	While solving them we take $N=5000$. In the mean field analysis the larger is $N$ the better are the results, but we find that the solutions of Eq.\ref{eq_r_a_with_real_part_drifting_osc_neglected} and Eq.\ref{eq_Omega} does not exhibit any visible change if we compare them for $N=1000$ and $N=5000$. The triangles in Fig. \ref{fig_analytical}(b) represents $r$ values for $N=1000$, while the continuous green line corresponds to $N=5000$. These solutions are indistinguishable and therefore we can safely take $N=5000$.
	
	Here we present the results from the mean-field analysis only for $\Delta \omega=1$ [Fig. \ref{fig_analytical}], while in Appendix \ref{appendix_b} we compare the mean-field predictions with the numerical simulations for $0 < \Delta \omega < 1$.  
	Although not shown in Fig. \ref{fig_analytical}, a globally synchronized state exists for all $\sigma$ values if $\alpha=0$. \cite{Kumar2020} Eq.~$9$ of Ref.~\onlinecite{Kumar2020} indicates that, at $\sigma=0$, the locked oscillator's natural frequencies satisfy the relation $\omega_i^a \le \lambda$; therefore, for any $\lambda \ge 0.5$, all the oscillators are locked. An increase in $\sigma$ will only brings the oscillators closer to each other; hence, the globally synchronized state exists for all $\sigma$ values. Considering $\alpha=0$ case and Fig. \ref{fig_analytical}, we conclude that an increase in $\alpha$ from $0$ to $\pi/2$ suppresses the synchronization among the oscillators and the globally synchronized state is converted to an incoherent state. In Figs. \ref{fig_analytical}(a-c), $r^a$ increases continuously from $0$ to $1$ by including more and more nodes in the locked state; a further increase in $\alpha$ turns the phase transition into a discontinuous one [Figs. \ref{fig_analytical}(d-f)].
	Interestingly, multiple $\psi^{a(b)}$ values (and therefore multiple $r$ values for the full multiplex network) can exist at $\alpha=\pi/2$ (Appendix \ref{appendix_a}), where $re^{i \psi} = (r^a e^{i \psi^a}+ r^b e^{i \psi^b})/2$. However, we do not find any notable difference in the corresponding $r^{a(b)}$ values.
	Around the bistable regime, the mean field analysis predicts three non-zero $r$ values in Figs. \ref{fig_analytical}(d),(e) and two non-zero $r$ values in Fig. \ref{fig_analytical}(f). The upper and lower $r$ values in Figs. \ref{fig_analytical}(d),(e) should represent the stable synchronized states, while those in the middle should represent an unstable state. \cite{Zhang2013, Leyva2013b, Kumar2020} We mention that in Fig.\ref{fig_analytical} we have neglected $r=\Omega=0$ solution of Eq.\ref{eq_r_a_with_real_part_drifting_osc_neglected} and Eq.\ref{eq_Omega} for $0 < \alpha \le \pi/2$; however, note that it can not be a solution if $\alpha=0$ (see Eq. $9$ of Ref.~\onlinecite{Kumar2020}).
	
	As illustrated by Fig. \ref{fig_analytical}, the mean-field results do not match with the numerical simulations.
	In the following, we justify these discrepancies.
	Firstly, at $\alpha=0.5$, Fig. \ref{fig_analytical}(a) show that the $r$ values for $\sigma \approx 0$ does not match with the mean-field predictions, which is due to the sensitivity of the $r$ values to the initial conditions and not due to failure of mean-field analysis. 
	Ignoring intralayer coupling terms for $\sigma \approx 0$, the frequency synchronization between the mirror nodes leads to the relation $sin(\theta_i^b-\theta_i^a) \approx (\omega_i^b-\omega_i^a) / 2 \lambda cos(\alpha)$. Therefore, synchronization among mirror nodes requires $\lvert{\omega_i^a}\rvert \le \lambda cos(\alpha) \approx 8.77$. Hence $\theta_i^b-\theta_i^a=\theta_i^{b^*}-\theta_i^{a^*}$ is constant (or almost constant) in time for all the oscillators and the phases in layer $a$ can be written as $\theta_i^a(t) \approx \int_{0}^{t} (\omega_i^a + \lambda sin(\theta_i^{b^*}-\theta_i^{a^*}+\alpha)) dt + \theta_i^a(t=0)$.
	This relation reveals that, depending on the initial phase distribution, different $r$ values can exist.
	
	Secondly, as $\alpha \rightarrow \pi/2$, $r$ values in the partially synchronized state do not match with the numerical simulations. 
	Also, if the unstable state is an indicator of the presence of hysteresis, the mean-field analysis does not project the position of the hysteresis correctly [Fig.\ref{fig_analytical}(d-f)]. 
	To find the reason behind the failure of the mean-field predictions, we examine the validity of the assumptions. For example, in the forward continuation we plot maximum fluctuation in $|sin(\psi^a-\psi^b)|$ after the initial transient time is passed. The stars in Figs. \ref{fig_analytical}(b-f) reveals that $\psi^a-\psi^b$ is not constant in the incoherent and partially synchronized regime. Therefore, the statements $\psi^a = \psi^b$ and $\dot \psi^{a(b)} =\Omega$ do not hold exactly, and a disagreement between the mean-field and the numerical results may occur.
	
	For $\alpha=0$ and for a small $\lambda$ value, Ref.~\onlinecite{Kumar2020} compares the mean-field predictions with numerical results. A good match there indicates that the mean-field analysis is valid for the small $\lambda$ values as well, as long as $\alpha=0$ or close to it.

	\section{An extension to the single-layer networks}
	\label{sec_6}
\begin{figure}[t]
		\centering
		\centerline{\includegraphics[width=\columnwidth]{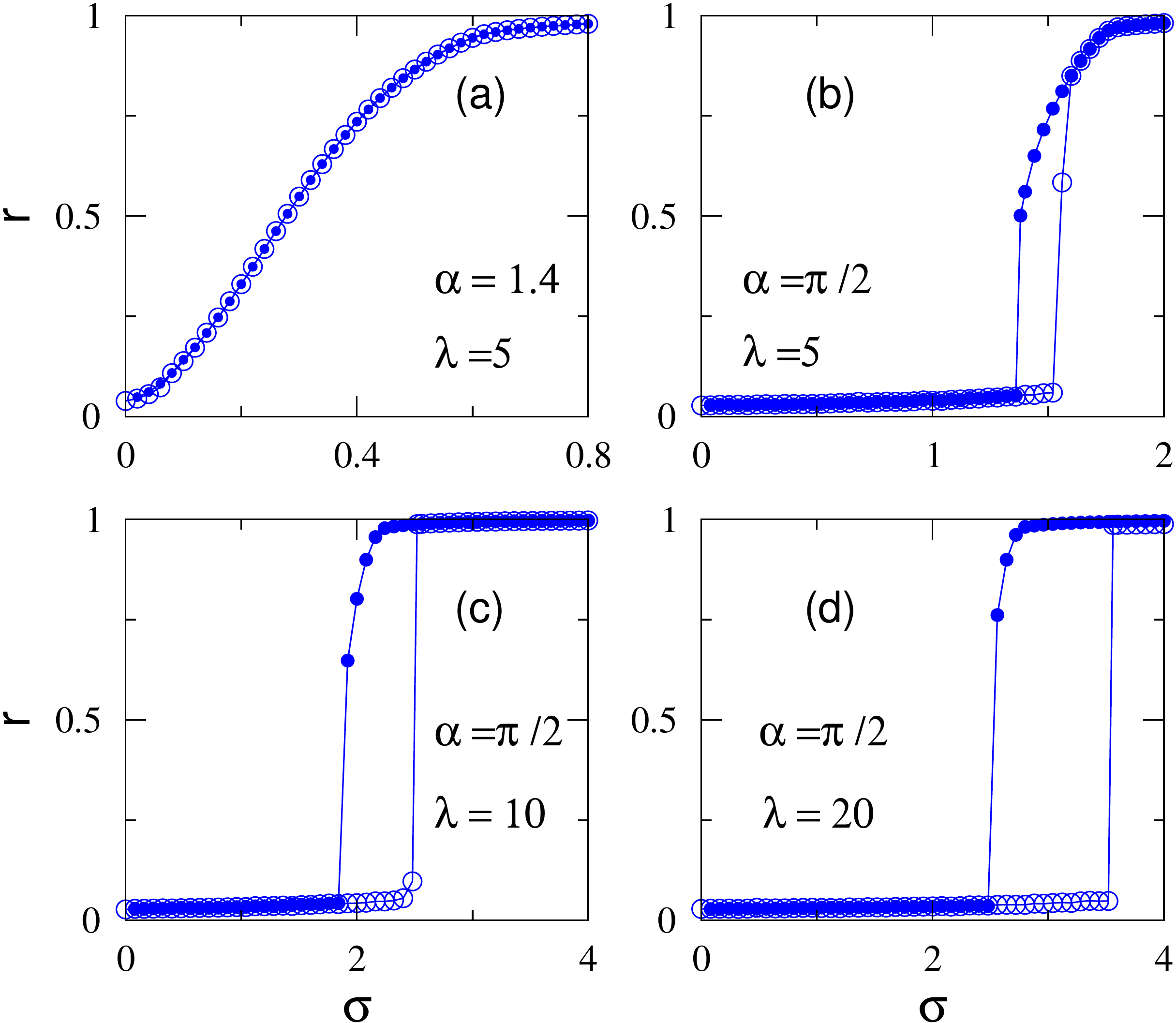}}
		\caption{(color online). (a,b) show the emergence of ES due to the phase shifted coupling in the single layer network given by Eq.\ref{eq_single_layer_model}. (c,d) shows a favourable impact of $\lambda$ on the hysteresis width. The network size (N) in all the plots is $1000$.} 
		\label{fig_single_network}
	\end{figure}

	As shown in Appendix \ref{appendix_b}, the phases in the synchronized state at $\Delta \omega=1$ satisfy the relation $\theta_i^a = \theta_{N-i+1}^b$, which reflects that a phase-shifted coupling between the opposite natural frequencies of a single layer network should also result in an ES. 
	Therefore, the model should be

	\begin{equation}
		\begin{split}
			\dot{\theta_i} = \omega_i + \frac{\sigma}{N} \sum_{j=1, j \neq N-i+1}^{N} \sin(\theta_j-\theta_i)+  \\
			\lambda \sin(\theta_{N-i+1}-\theta_i + \alpha),
		\end{split}
		\label{eq_single_layer_model}
	\end{equation}

	where $\omega_i = -\omega_{N-i+1}$ and $i=1,2 . . . N$. Eq. \ref{eq_single_layer_model}, infact, represents a globally connected network of heterogeneous coupling. Fig. \ref{fig_single_network} reveals that, indeed, the model represented by Eq.\ref{eq_single_layer_model} exhibit ES along with a hysteresis. Figs. \ref{fig_single_network}(c,d) further illustrates that, similar to the multiplex networks, an increase in $\lambda$ has a favourable impact on the hysteresis width. 
	Other choices of natural frequencies for the phase-shifted coupling and their impact on ES may be explored further elsewhere. The obtained results expand the scope of the phase-shifted coupling in obtaining ES to the basic framework of the networks.

	\section{Conclusion}
	Different methods have been proposed in the past years to incite ES in the coupled phase oscillators.
	In this work, we have demonstrated that a phase shift in the interlayer coupling terms of a multiplex network of Kuramoto oscillators can also trigger ES in the layers. As $\alpha \rightarrow \pi/2$, ES emerges along with a hysteresis. The hysteresis width can be controlled by the phase shift value, the interlayer coupling strength, and the frequency mismatch between the mirror nodes. Similar to earlier studies, we find that the suppression of synchronization is the underlying mechanism behind the onset of ES. We have tested the robustness of ES against changes in the network topology and frequency distribution. Homogeneous random networks manifest a significant jump in the phase transition, while for the heterogeneous networks the jump size is negligible. 
	A strong discontinuity in the phase transition for globally connected layers suggests that an increase in the average connectivity of the layers should have a favourable impact on the jump size, and therefore heterogeneous layers may exhibit a sizeable jump for sufficient high average connectivity.
	We have performed a mean-field analysis, which shows a fair agreement with the numerical results if $\alpha$ values remain close to $0$.
	Finally, we have extended the results to the single-layer networks validating that the basic framework of the networks is within the reach of the phase-shifted coupling to get ES.
	Also, note that the removal of a small fraction of interlayer links should not have much impact on the ES; therefore, the results from our work should be relevant to the more general multilayer networks.
	
	Within the single layer framework, the phase-shifted couplings in the Kuramoto oscillators have attracted significant attention; we expect it to exist in the real-world multilayer systems as well. \cite{Maximilian2015, Bastian2019}

	\acknowledgments 
	SJ acknowledges the Government of India, CSIR grant 25(0293)/18/EMR-II and DST grant EMR/2016/001921 for financial support. AK acknowledges CSIR, the Government of India for providing SRF.

	\appendix
	\section{$\theta_i^b-\theta_i^a$ values in the synchronized state}
	\label{appendix_a}
	Here onwards, we restrict the mean-field analysis only for $0 \le \alpha \le \pi/2$; however, as explained later it can be applied to any $\alpha$ value. For $0 \le\alpha < \pi/2$, we further assume that $\psi^a=\psi^b$; while for $\alpha=\pi/2$, other relations between $\psi^a$ and $\psi^b$ can also exist.
	With $\psi^a=\psi^b$, Eq.\ref{eq_theta_a_minus_theta_b} becomes

	\begin{align}
		&{(\omega_i^{b}-\omega_i^{a}- 2 \lambda cos(\alpha) \sin(\theta_i^{b}-\theta_i^{a}))}^2 \cos^2{(\frac{\theta_i^{b}-\theta_i^{a}}{2})} + \nonumber\\
		&{(2\Omega -(\omega_i^{b}+\omega_i^{a}) - 2 \lambda \sin(\alpha) \cos(\theta_i^{b}-\theta_i^{a}))}^2 \sin^2{(\frac{\theta_i^{b}-\theta_i^{a}}{2})} \nonumber\\
		&= \sigma^2 r^2 \sin^2{({\theta_i^{b}-\theta_i^{a}})}.
		\label{eq_theta_a_minus_theta_b_with_psi_a_equal_to_psi_b}
	\end{align}
	
	In general, Eq. \ref{eq_theta_a_minus_theta_b_with_psi_a_equal_to_psi_b} can be written as a sixth order polynomial in $\cos({\theta_i^{b}-\theta_i^{a}})$, while at $\alpha = \pi/2$ it is a third order polynomial in $\cos({\theta_i^{b}-\theta_i^{a}})$. To our knowledge, the roots of the sixth order polynomial are not known in terms of a formula; thus, we find its roots numerically. Out of the multiple roots, a physically accepted root is selected as follows. 
	For finite $\sigma$ values, the L.H.S. of Eq.\ref{eq_theta_a_minus_theta_b_with_psi_a_equal_to_psi_b} is positive at $\theta_i^b -\theta_i^a =0, \pm \pi$, suggesting that Eq.~\ref{eq_theta_a_minus_theta_b_with_psi_a_equal_to_psi_b} has even number of roots in the range $0, \pm \pi$. Numerically, it can be checked that Eq.~\ref{eq_theta_a_minus_theta_b_with_psi_a_equal_to_psi_b} has two real roots in the range $0$ to $\pm \pi$. 
	With $\lambda$ being fixed, dividing Eq.\ref{eq_theta_a_minus_theta_b_with_psi_a_equal_to_psi_b} by $\sigma^2$ and increasing it infinitesimally reflects that $\theta_i^b-\theta_i^a \rightarrow 0, \pm \pi$; therefore, the roots closer to $0$ approach towards it. Here we have taken into account the fact that $\Omega$ (Eq. \ref{eq_Omega}) remains bounded as $\sigma$ increases. 
	Since an increase in $\sigma$ should brings the phases in layer $a(b)$ closer to $\psi$ and therefore $\theta_i^b-\theta_i^a$ should decrease, only the roots closer to $0$ should be accepted. Next, for $\omega_i^b -\omega_i^a \gtrless 0$, we only take a root such that $\theta_i^b - \theta_i^a \gtrless 0$, which can be justified easily from Eq. \ref{eq_sum_and_difference}(b). It can be re-written as
	
	\begin{equation}
		\begin{split}
			2 \sin{\big(\frac{\theta_i^{b}-\theta_i^{a}}{2}\big)} \cos{\big(\frac{2 \psi - \theta_i^{a} -\theta_i^{b}}{2}\big)} + \\
			{\frac{2 \lambda cos (\alpha) sin\big(\theta_i^{b}-\theta_i^{a}\big)}{\sigma r}} = \frac{\omega_i^b - \omega_i^a}{\sigma r}.
			\label{eq_to_short_root}
		\end{split}
	\end{equation}

	Again, as $\sigma \rightarrow \infty$, Eq.\ref{eq_to_short_root} indicates that either $\theta_i^{b} - \theta_i^{a} \rightarrow 0$ or $(2\psi - \theta_i^{a} - \theta_i^{b}) \rightarrow  \pm \pi$. 
	In the first case the roots approach to $0$ such that $\theta_i^b - \theta_i^a \gtrless 0$ for $\omega_i^b -\omega_i^a \gtrless 0$. Therefore, out of the four roots, a physically accepted root is selected by using these arguments.
	
	Note that, if $\Delta \omega =0$, $\theta_i^b-\theta_i^a=0$ is a root of Eq.\ref{eq_theta_a_minus_theta_b_with_psi_a_equal_to_psi_b}. 
	The identical phases of the mirror nodes also justify the impact of $\Delta \omega$ on ES [Figs \ref{fig_hysteresis_size}(d-f)]. 
	The interlayer coupling terms in this case becomes $\lambda \sin(\alpha)$. Therefore, the addition of a constant term to all the oscillators makes the synchronized state for the multiplex network the same as an isolated network. Since an isolated globally connected network with uniform frequency distribution exhibit a discontinuous phase transition without any hysteresis. \cite{Pazo2005} we can expect no hysteresis with a decrease in $\Delta \omega$.

	At $\alpha=\pi/2$, as mentioned above Eq.~\ref{eq_theta_a_minus_theta_b_with_psi_a_equal_to_psi_b} turns into a third order polynomial 
	$x_i^3 + p_ix_i^2+q_ix_i+r_i' =0$, where $x_i=\cos(\theta_i^b -\theta_i^a)$. The coefficients of the polynomial are given by
	
	\begin{align*}
		\begin{split}	
			p_i &= \frac{-\sigma^2 r^2 - 2 \lambda^2 + 2 \lambda \{\omega_i^a+\omega_i^b -2 \Omega\}} {2 \lambda^2}, \nonumber
		\end{split}\\
		\begin{split}
			q_i &=\frac{ - (\omega^b-\omega^a)^2 - 4 \lambda  \{\omega_i^a+\omega_i^b -2 \Omega\}} { 4 \lambda^2} \nonumber \\ &+ \frac{ (\omega_i^a+\omega_i^b -2 \Omega)^2} {4 \lambda^2},
		\end{split}\\
		\begin{split}
			r_i' &= \frac{2 \sigma^2 r^2 - (\omega_i^b-\omega_i^a)^2 - (\omega_i^a + \omega_i^b -2 \Omega)^2}{4 \lambda^2} \nonumber.
		\end{split}\\
		\label{}
	\end{align*}
	
	And the roots of the polynomial are: \cite{roots}
	\begin{align}
		&x_{i_1}= U_i+ V_i -\frac{p_i}{3},\\
		&x_{i_2}= -\frac{(U_i+V_i)}{2} + i\frac{\sqrt{3}}{2} \frac{(U_i-V_i)}{2} -\frac{p_i}{3}, \\ 
		&x_{i_3}= -\frac{(U_i+V_i)}{2} - i\frac{\sqrt{3}}{2} \frac{(U_i-V_i)}{2} -\frac{p_i}{3}, 
		\label{root_third_order_pol}
	\end{align}
	
	where
	\begin{align}
		U_i&= \Big\{\frac{-b_i}{2} + \sqrt{\frac{b_i^2}{4}+ \frac{a_i^3}{27}}\Big\}^{1/3}, \nonumber \\
		V_i&= \Big\{\frac{-b_i}{2} - \sqrt{\frac{b_i^2}{4}+ \frac{a_i^3}{27}}\Big\}^{1/3}, \label{u_v_a_b_values}\\
		a_i&= (3 q_i - p_i^2)/3,\nonumber\\
		b_i&= (2 p_i^3 - 9 p_i q_i + 27 r_i')/27. \nonumber \\ \nonumber
	\end{align}

	By calculating $x_i$ values numerically, we can find that the physically accepted root is $x_{i_3}$. For example, taking $\Delta \omega=1$, $\omega_i^a =-0.5$, and $\lambda=1$, we calculate $x_i$ values at $\sigma =2, 20$. As $\sigma \rightarrow \infty$, $r \rightarrow 1$ and $\Omega \rightarrow \lambda$; therefore, $r ,\Omega$ can be taken as $1, \lambda$. We find that at $\sigma=2$, $x_{i_1}, x_{i_2}, x_{i_3}$ are approximately $4.32, -0.19, 0.87$, while at $\sigma=20$ they are approximately $202, -0.98, 0.99$, respectively. The root $x_{i_1}$ is greater than $1$ and therefore it is physically un-acceptable. Out of the remaining two roots $x_{i_2}$ approaches towards $\pi$; therefore, it can also not be the root.
	
	We mention that a similar mean-field analysis can be performed for $\pi/2 < \alpha \le 2 \pi$. If $\pi/2 < \alpha < 3 \pi/2$, the coefficient of $sin(\theta_i^{a(b)}-\theta_i^{b(a)})$ is negative and therefore $\psi^a-\psi^b \neq 0$. $\Delta \omega=0$ case suggest that $\theta_i^b-\theta_i^a =\pm \pi$, $\psi^a-\psi^b =\pm \pi$ (Eq.\ref{eq_theta_a_minus_theta_b}). Similarly, for $0<\Delta \omega \le 1$, by following a parllel analysis as in Appandix \ref{appendix_b}, one can check that $\psi^a-\psi^b \simeq \pi$. Also, the roots of Eq.\ref{eq_theta_a_minus_theta_b} which are closer to $\pm \pi$ should be selected. Furthermore, due to same sign of $\sin(\theta_i^{a(b)}-\theta_i^{b(a)})$ terms, the mean-field analysis for $3 \pi/2 < \alpha \le 2 \pi$ should remain same as for $0 \le \alpha < \pi/2$.

	\subsection{Multi-stability at $\alpha = \frac{\pi}{2}$ and $\frac{3\pi}{2}$}

	We find that, besides $\psi^a=\psi^b$, other relations between $\psi^a$ and $\psi^b$ can also exist if $\alpha$ is $\pi/2$ or $3 \pi/2$.
	Multiple $\psi^a - \psi^b$ values at $\alpha = \pi/2$ can exist since there is no attractive coupling (sine coupling) between the layers which can decide the interlayer position of the oscillators. 	
	Eq.\ref{eq_model} claims that the phases in layer $a(b)$ must satisfy the relation
	\begin{equation}
		\begin{split}
			\sum_{j=1}^N {(\theta_i^a -\theta_i^b)}_t = \sum_{j=1}^N (\theta_i^a -\theta_i^b)_{t=0}=k.
			\label{eq_phase_criteria_for_alpha_pi_by_2} 
		\end{split}
	\end{equation}
	In this work we take $k=0$, but an addition of $2n\pi$ to $k$, where $n=\pm 1, \pm 2$, does not make any difference physically.
	Next, the synchronized state (Eq.\ref{eq_synchronized_state}) at $\alpha=\pi/2$ can be written as
	
	\begin{equation}
		\begin{split}
			\theta_i^{a(b)}= \psi^{a(b)} - asin(\frac{\Omega-\omega_i^{a(b)} \mp \lambda cos(\theta_i^{b(a)} -\theta_i^{b(a)})}{\sigma r}.)
		\end{split}
		\label{eq_phases_in_synch_state_alpha_pi_by_2}
	\end{equation}
	Here $\mp$ corresponds to $\alpha=\frac{\pi}{2}, \frac{3\pi}{2}$, respectively.
	First, we prove the existance of multiple synchronized states for $\Delta \omega=0, 1$. In the first case, Eq.\ref{eq_theta_a_minus_theta_b} shows that $\theta_i^b-\theta_i^a = \psi^b-\psi^a$ is a root. 
	Putting $\theta_i^b-\theta_i^a$ values in Eq.\ref{eq_Omega}, it returns $\Omega = \lambda cos(\psi^b-\psi^a)$; therefore, Eq.\ref{eq_phases_in_synch_state_alpha_pi_by_2} becomes 
	
	\begin{equation}
		\begin{split}
			\theta_i^{a(b)}= \psi^{a(b)} + asin(\frac{\omega_i}{\sigma r}),
		\end{split}
		\label{eq_phases_synch_state_del_alpha_pi_by_2_del_omega_0}
	\end{equation}
	
	where $\omega_i=\omega_i^a=\omega_i^b$. After putting $k=2n\pi$ and $\theta_i^{a(b)}$ values in Eq.\ref{eq_phase_criteria_for_alpha_pi_by_2}, we get $\psi^a -\psi^b = 2 n\pi/N$.
	
	For $\Delta \omega=1$, the natural frequencies satisfy the relation $\omega_i^a =\omega_{N-i+1}^b$, where $i=1, 2 . . . N$. Therefore, from Eq.\ref{eq_theta_a_minus_theta_b}, it can be easily perceived that irrespective of the $\alpha$ value we have $\theta_i^b -\theta_{i}^a = \theta_{N-i+1}^a -\theta_{N-i+1}^b$. 
	\begin{figure}[t]
		\centering
		\centerline{\includegraphics[width=0.9\columnwidth]{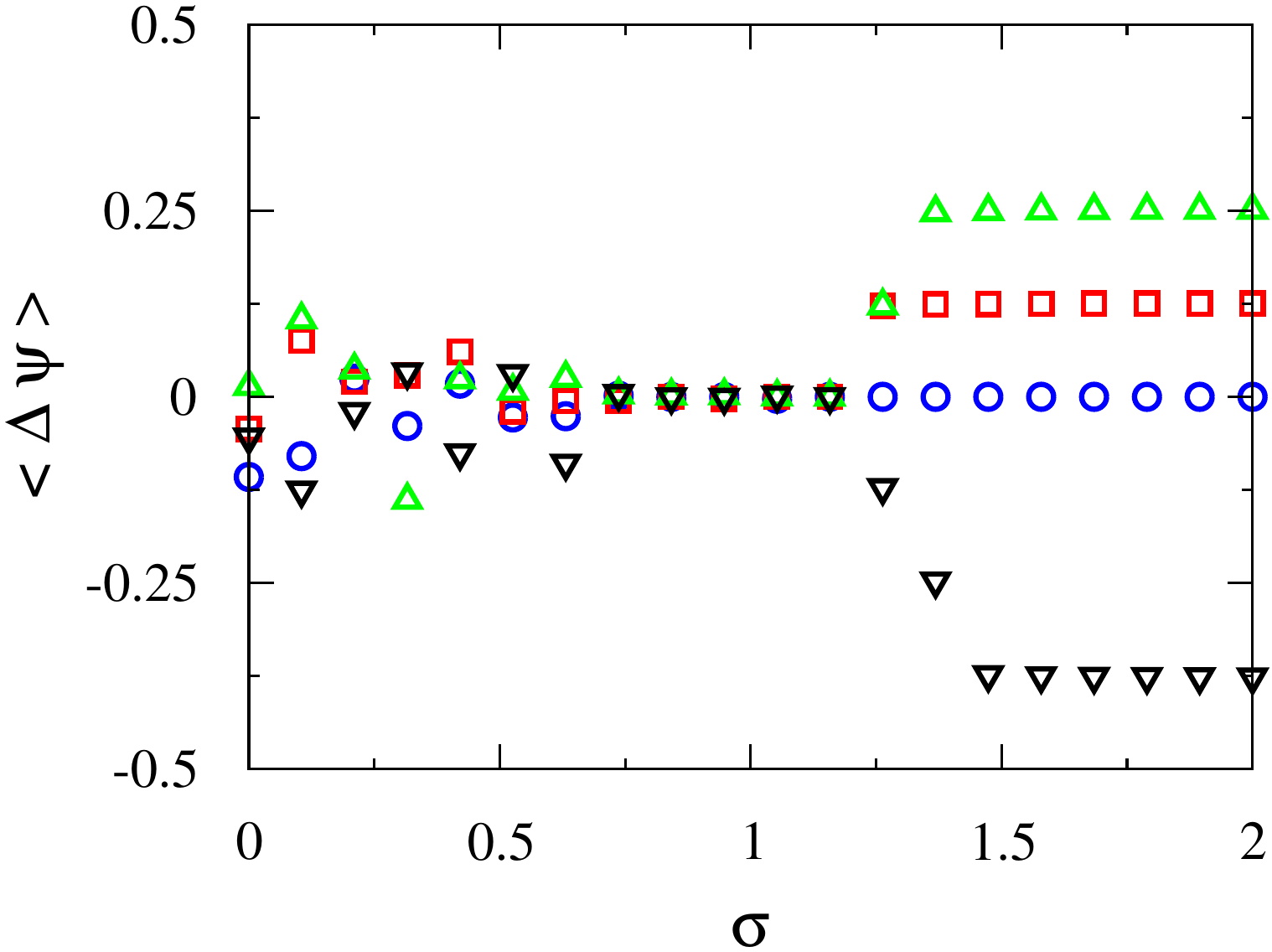}}
		\caption{(color online). The time averaged $\Delta \psi =\psi^a -\psi^b$ values manifest multiple synchronized states. 
			The circles, squres, upper-triangles, and lower-triangles correspond to $\Delta \psi=0, \frac{\pi}{25}, \frac{2 \pi}{25}, \frac{-3 \pi}{25}$, respectively. The layers are globally connected with $N=50$, $\lambda=1$, $\alpha=\pi/2$, and $\Delta \omega =1$.} 
		\label{fig_multistability}
	\end{figure}
	It makes the second term in the R.H.S of Eq.\ref{eq_phases_in_synch_state_alpha_pi_by_2} same for oscillators of natural frequencies $\omega_i^a$ and $\omega_{N-i+1}^b$. With this fact again putting $\theta_i^{a(b)}$ values and $k$ in Eq.\ref{eq_phase_criteria_for_alpha_pi_by_2} we get $\psi^a -\psi^b = 2n\pi/N$. Fig. \ref{fig_multistability} reports some of these states corresponding to $n=0,1,2,-3$ from the numerical simulations.
	
	For $0 < \Delta \omega < 1$ also, the same relation between $\psi^{a(b)}$ can be derived in the limit $\sigma \rightarrow \infty$. As $\sigma$ increases, we know that $\theta_i^{a(b)}\rightarrow \psi^{a(b)}$; therefore, $\theta_i^b-\theta_i^a \rightarrow \psi^b-\psi^a$, which is same as we found for $\Delta \omega=0$. Hence, for large $\sigma$ values we get the same relation between $\psi^{a}$ and $\psi^{b}$ for any $\Delta \omega$ value.
	
	\subsection{Linear Stability analysis}
	Using the linear stability analysis, we prove the stability of the multiple states in the limit $\sigma \rightarrow \infty$.
	The intralayer entries of the Jacobian are such that $J_{ij} =(\sigma/N) cos(\theta_j-\theta_i)$, where $i,j=1,2, . . .,N$ for links in layer $a$ and $i,j=N+1, N+2, . . .,2N$ for links in layer $b$; the interlayer entries are such that $J_{i, i+N} = - J_{i+N, i} = -\lambda sin(\theta_{i+N}-\theta_i)$ and $0$ otherwise. And finally, the diagonal entries are $J_{ii} =-\sum_{j=1}^{2N} J_{ij}$, where $i=1,2,3 \; . \; \; . \;\; . \;\; 2N$. Now, as $\sigma \rightarrow \infty$, intra, interlayer entries approach $\sigma/N, \, \pm \lambda \sin(\psi^b-\psi^a)$, respectively.  
	Therefore, it can be easily checked that the eigenvalues of the Jacobian matrix are $0, 0 , -\sigma, -\sigma, . . ., -\sigma$. All the non-zero eigenvalues are negative, and hence the multiple synchronous states are linearly stable. \cite{Strogatz2000} Although the eigenvalue $0$ appears twice, its corresponding eigenvector's entries should all be $1$; therefore, the degeneracy in it does not affect the stability. \cite{Strogatz2000} Independence of the eigenvalues from the interlayer entries is perhaps due to the negative signs of the interlayer entries of the Jacobian, which makes the trace independent of the interlayer entries, hinting that the eigenvalues should not depend on the orientation of $\psi^a$ and $\psi^b$.

	\section{Validity of the assumptions}
	\label{appendix_b}
	Except the time dependence of $r^{a(b)}$ or $\psi^{a(b)}$ as $\alpha \rightarrow \pi/2$, the assumptions (a) $r^a=r^b$ and (b) $\psi^a=\psi^b$ are valid. 
	First, we justify these assumptions for $\Delta \omega=0, 1$. 
	Also, we discuss the cases $0 \le \alpha < \pi/2$ and $\alpha=\pi/2$ separately. 
	For $\Delta \omega =0$ the phases of the mirror nodes are identical (Appendix \ref{appendix_a}) and therefore the assumptions (a), (b) are valid from Eq.\ref{eq_r_a_b}. Along with $\psi^a=\psi^b$, the earlier discussion for $\Delta \omega=1$ and $\alpha=\pi/2$ (Appendix \ref{appendix_a}) will return us $\theta_i^b =\theta_{N-i+1}^a$. Since the distribution of phases is identical for layer $a(b)$, the assumptions are valid. 
	Finally, the only assumption (a) is valid at $\alpha=\pi/2$. Eq.~\ref{eq_phases_synch_state_del_alpha_pi_by_2_del_omega_0} implies that $\psi^a-\theta_i^a=\psi^b-\theta_{i}^b$ (for $\Delta \omega=0$), while Eq.\ref{eq_synchronized_state}(a),(b) show that $\psi^a-\theta_i^a=\psi^b-\theta_{N-i+1}^b$ (for $\Delta \omega=1$); therefore, Eq.\ref{eq_r_a_b} returns $r^a=r^b$.

	\begin{figure}[t]
		\centering
		\centerline{\includegraphics[width=0.9\columnwidth]{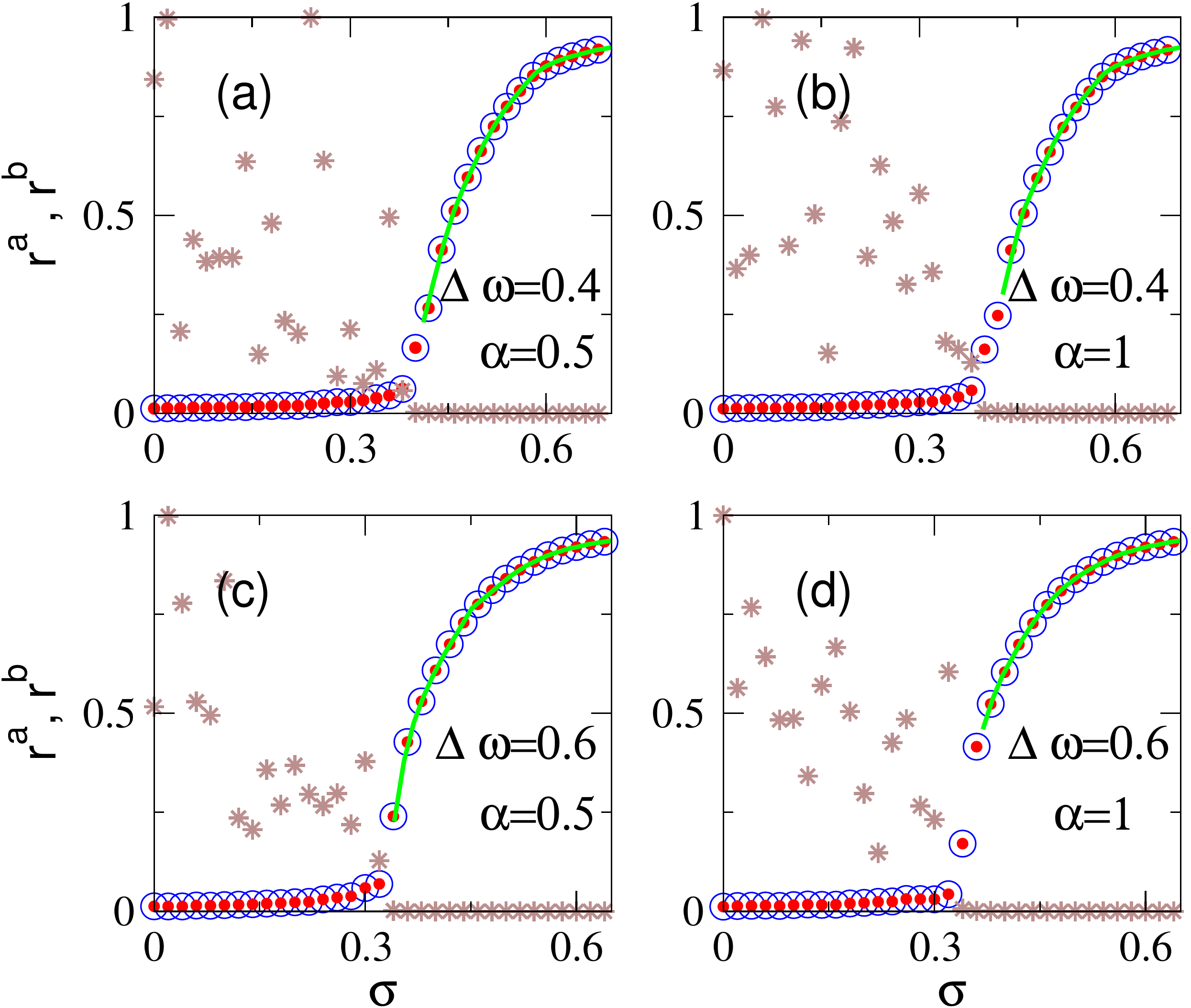}}
		\caption{(color online). (a-d) show a comparison of $r^a$ (open blue circles) and $r^b$ (filled red circles) values taken from the numerical simulations. The stars represent maximum fluctuation in $|sin(\psi^a-\psi^b)|$ after the initial transient time is passed. 
			The continuous green lines represent $r^a$ values from the mean-field analysis.  
			The layers are globally connected with $N=5000$ and $\lambda=10$.} 
		\label{fig_ana_num_del_neq_1}
	\end{figure}

	For $0 < \Delta \omega < 1$, with the help of the numerical simulations and by establishing a good match between the numerical and mean field predicted $r$ values, we show that the assumptions (a), (b), and (c) the exclusion of the drifting oscillators in Eq.\ref{eq_r_a_with_real_part} and Eq.\ref{eq_Omega} are valid. Since $r^{a(b)}$, $\psi^{a(b)}$ starts depending on time as $\alpha \rightarrow \pi/2$, here we prove the validity of the assumptions only for $\alpha$ values far from $\pi/2$. For $0 \le \alpha < \pi/2$, note that the assumptions (a), (b) are motivated from the fact that an increase in $\lambda$ increases the $sine$ coupling between the mirror nodes which will causes $\theta_i^b \rightarrow \theta_i^a$ and therefore $r^a \rightarrow r^b$ and $\psi^a \rightarrow \psi^b$. 
	
	Figs. \ref{fig_ana_num_del_neq_1}(a-d) illustrates that almost same $r^{a(b)}$ values from the numerical calculations validate the assumption (a). Furthermore, the analytical predicted $r$ values are also in excillent agreement with the numerical calculations. The stars indicates that the assumption (b) also holds in the synchronized regime.
	
	For the assumption (c), only based on the excellent agreement between the numerical and the mean-field predicted $r$ values, we claim that the exclusion of the drifting oscillators is a fair choice. Also, note that the drifting oscillators decrease as $r \rightarrow 1$ and therefore the assumption (c) holds automatically for larger $r$ values.
	
	As displayed by Fig.~\ref{fig_ana_num_del_neq_1}, Eq.\ref{eq_r_a_with_real_part} and Eq.\ref{eq_Omega} does not have any solution for small $r$ values. Non existence of any solution is possibly due to failure of the assumptions (a),(b) for small number of locked oscillators. 
	For example, if we compare the RHS in Eq.\ref{eq_Omega} for layer $a$ and $b$, the terms $\sum \omega_i$ and $\sum \sin(\theta_i^b -\theta_i^a)$ may be not be same for a very small number of locked oscillators and for a finite $N$, indicating the failure of the assumption (a) or (b).
	
	Moreover, we find that the RHS of Eq.\ref{eq_r_a_with_real_part} and Eq.\ref{eq_Omega} are not a smooth functions for small $r$ values and therefore multiple solutions of Eq.\ref{eq_r_a_with_real_part} and Eq.\ref{eq_Omega} can exist; while plotting Fig.\ref{fig_ana_num_del_neq_1}, we have selected only the largest of the multiple solutions for $0 < r \lesssim 0.3$.
	Remember that for $\Delta \omega=0,1$ the R.H.S. of Eq.\ref{eq_r_a_with_real_part} and Eq.\ref{eq_Omega} are smooth for all $r$ values because the locked oscillators adds to the synchronized state systematically i.e. smaller natural frequencies synchronise first followed by the larger ones. 
	It is due to this reason we primarily focused on the mean-field analysis only for $\Delta \omega=1$.

	\bibliographystyle{apsrev4-1}
	\bibliography{references_multiplex_2.bib}

\begin{thebibliography}{44}
\bibitem [{rk4()}]{rk4_method}%
  \BibitemOpen
  \href@noop {} {}\bibinfo {note} {We use ode45 solver in MATLAB. Although the
  relative and absolute error tolerences differ in different numerical
  simulaitons, their maximum values are $10^-7$, $10^-6$,
  respectively.}\BibitemShut {Stop}%
\bibitem [{Lor()}]{Lorentz_frequencies}%
  \BibitemOpen
  \href@noop {} {}\bibinfo {note} {The Lorentzian frequencies are generated
  using inverse transform sampling i.e. using the formula
  $(1/2)tan(\pi(w-0.5))$. Here $w$ is a vector containing uniformly distributed
  numbers $w_i=(i-1)/(N-1)$, with $i=2, 3 . . .N-1$.}\BibitemShut {Stop}%
\end{thebibliography}

\end{document}